\documentstyle[aps,preprint]{revtex}
\begin{document}
\draft

\title{Glass Formation in the Gay-Berne nematic liquid crystal}

\author{A. M. Smondyrev and Robert A. Pelcovits}

\address{ Department of Physics, Brown University, Providence RI, 02912}

\date{\today}
\maketitle

\begin{abstract}
We present the results of molecular dynamics simulations of the Gay-Berne model of liquid crystals, supercooled from the nematic phase. We find a glass transition to a metastable phase with nematic order and frozen translational and orientational degrees of freedom.  For fast quench rates the local structure is nematic-like, while for slower quench rates smectic order is present as well.  
 \\
\end{abstract}

\narrowtext
%\twocolumn
The study of nematic glasses \cite {expt} is richer than the corresponding study of isotropic glasses due to the presence of orientational degrees of freedom in the former systems. Assuming that we supercool a liquid crystal starting from its nematic phase (rather than the isotropic phase) we expect a nematic glass to have long-range orientational order like an equilibrated nematic as well as frozen density and director fluctuations.  Orientationally this glassy phase is similar to a mixed magnetic phase where both ferromagnetic and vector spin glass order coexist \cite{SG}. However, unlike a spin glass the nematic glass is a nonequilibrium phase of matter and the molecular translational degrees of freedom freeze as well at the glass transition. 

In this Letter we study the formation of a nematic glass using molecular dynamics (MD). We model the liquid crystal using the Gay-Berne (GB) potential \cite{GB} which is an anisotropic Lennard-Jones potential. Previous molecular dynamics studies have indicated that the Gay-Berne potential exhibits a rich phase diagram \cite{deM,Luck90,LS93} and the principal dynamical features  \cite{DeRull,Sarman1,Sarman2,Smondyrev} of real liquid crystals.
 The GB potential is given by,
\begin{eqnarray}
\label{eq:U}
U({\bf \hat u}_1, {\bf \hat u}_2, {\bf r})&= & 4\varepsilon ({\bf \hat u}_1,
{\bf \hat u}_2, {\bf r})
\times \biggl[\biggl\{{\sigma_o\over r-\sigma({\bf \hat u}_1, {\bf \hat u}_2,
{\bf r}) +\sigma_o}\biggr\}^{12} \nonumber\\
& &-\biggl\{{\sigma_o\over r-\sigma({\bf \hat u}_1, {\bf \hat u}_2, {\bf
r})+\sigma_o}\biggr\}^6\biggr]
\end{eqnarray}
where ${\bf \hat u}_1, {\bf \hat u}_2$ are unit vectors giving the orientations
of the two molecules separated by the position vector $\bf r$. The parameters
$\varepsilon({\bf \hat u}_1, {\bf \hat u}_2, {\bf r})$ and $\sigma({\bf \hat
u}_1, {\bf \hat u}_2, {\bf r})$ are orientation dependent and give the well
depth and the intermolecular separation where $U=0$ respectively.
The well depth is written as,
\begin{equation}
\label{eq:epsr}
\varepsilon({\bf \hat u}_1, {\bf \hat u}_2, {\bf r})=\varepsilon_o
\varepsilon^\nu ({\bf \hat u}_1, {\bf \hat u}_2)\varepsilon^{\prime\mu} ({\bf
\hat u}_1, {\bf \hat u}_2, {\bf r})
\end{equation}
where
\begin{equation}
\label{eq:epsnor}
\varepsilon({\bf \hat u}_1, {\bf \hat u}_2)=(1-\chi^2 ({\bf \hat u}_1 \cdot{\bf
\hat u}_2)^2)^{-1/2}
\end{equation}
and
\begin{eqnarray}
\label{eq:epsprime}
\varepsilon^\prime ({\bf \hat u}_1, {\bf \hat u}_2, {\bf \hat r})&= &
1-{\chi^\prime\over 2}\biggl\{ {({\bf \hat r}\cdot{\bf u}_1
+{\bf \hat r}\cdot{\bf u}_2)^2\over 1+\chi^\prime
({\bf u}_1\cdot {\bf u}_2)} \nonumber\\
& &+ {({\bf \hat r}\cdot {\bf u}_1-{\bf \hat r}\cdot{\bf u}_2)^2\over
1-\chi^\prime({\bf u}_1\cdot{\bf u}_2)}\biggr\}
\end{eqnarray}
The range parameter $\sigma({\bf \hat
u}_1, {\bf \hat u}_2, {\bf r})$ is given by
\begin{eqnarray}
\sigma ({\bf \hat u}_1, {\bf \hat u}_2, {\bf \hat r})&= & \sigma_o\Biggl\lbrace
1-{\chi\over 2}\biggl\{ {({\bf \hat r}\cdot{\bf u}_1
+{\bf \hat r}\cdot{\bf u}_2)^2\over 1+\chi
({\bf u}_1\cdot {\bf u}_2)} \nonumber\\
& &+ {({\bf \hat r}\cdot {\bf u}_1-{\bf \hat r}\cdot{\bf u}_2)^2\over
1-\chi({\bf u}_1\cdot{\bf u}_2)}\biggr\}\Biggr\rbrace^{-1/2}
\end{eqnarray}
The shape anisotropy parameter $\chi$ is given by
\begin{equation}
\label{eq:chi}
\chi=\{(\sigma_e/\sigma_s)^2 -1\}/\{({\sigma_e/\sigma_s})^2+1\}
\end{equation}
where $\sigma_e$ and $\sigma_s$ are the separation of end-to-end and
side-by-side molecules respectively. The parameter $\chi^\prime$ is given by
\begin{equation}
\label{eq:chipr}
\chi^\prime
%% FOLLOWING LINE CANNOT BE BROKEN BEFORE 80 CHAR
=\{1-(\varepsilon_e/\varepsilon_s)^{1/\mu}\}/\{1+(\varepsilon_e/\varepsilon_s)^{1/\mu}\}
\end{equation}
The ratio of the well depths for end-to-end and side-by-side  configurations
 is $\varepsilon_e/\varepsilon_s$.

We investigated glass formation in the Gay-Berne fluid using a constant-pressure, 
constant-temperature MD method \cite{NH}, which allows the volume to change as a function of 
the temperature and pressure of the system. The edges of the cell were allowed 
to vary independently, but the orthogonal shape was maintained. We simulated a system of $N=864$ particles.  We chose $ \sigma_{e} / \sigma_{s} = 3,  \epsilon_{e} / \epsilon_{s} = 5 $, and
$\nu = 1$ and $\mu = 2$ as in the original work of Gay and Berne.
The moment of inertia was chosen to be $4 m \sigma_o^2$,
as in ref. \cite{Smondyrev}. We used periodic boundary conditions, and cut off and
smoothed the potential at  $ 3.8 \sigma_{0} $. The equations of motion were
solved
using the leap-frog algorithm with an integration time-step
$ \Delta t^{*} = 0.001 $ in dimensionless units
 ($\Delta t^{*} = \Delta t ( m \sigma_{0}^{2} / \epsilon_{0} )
^{-1/2}$, where $m$ is the mass of a molecule ).  The
initial configuration was chosen in the nematic phase at the dimensionless 
temperature $T^{*}(\equiv k_B T / \epsilon_0)=1.2$ and pressure $P^{*}(\equiv P \sigma_0^3 / \epsilon_0)=5.8$ \cite{Pressure}. The nematic order parameter
at this point was $S=0.75$. The temperature was then reduced in a sequential
fashion allowing the system to relax at each temperature for a number of  time
steps depending on the quench rate. The quench rate is defined by the ratio of the change in dimensionless temperature
to the number of timesteps between two succeeding temperature reductions.  Our discussion will
focus primarily on simulations where the temperature was
lowered in decrements of $\Delta T^{*} = 0.1$ every 100 iterations, corresponding in real units to a quench rate approximately equal to $10^{13}$ K/s. Rates faster and slower than this one will be discussed at the end. After the system was cooled to its final temperature, it was allowed to anneal and various structural and dynamical properties were measured. 
 Fig.~\ref{fig1} shows the potential energy 
as a function of the annealing time at four different final temperatures. The breaks in 
the curves corresponding to temperatures $T^{*}=0.4$ and $T^{*}=0.5$ suggest that  
crystallization may have occurred. Further proof of crystallization can be found 
by examining pair distribution functions after the breaks take place.
In Fig.~\ref{fig2} we show  parallel, perpendicular (relative to the director) and orientationally--averaged pair distribution functions at $T^{*}=0.4$ after an annealing run of 120,000 iterations.  From the plot of the parallel distribution function we see
that a smectic density wave has been formed parallel to the director. The 
peaks of the perpendicular distribution function show the hexagonal crystalline ordering within the layers.
The splitting of the second peak of the orientationally--averaged distribution function is typical of a crystalline phase.  The nonzero value of the perpendicular distribution function (which is averaged over the layers) at $r=0$ shows  the strong correlation between the layers. At $T^{*}=0.3$ and below, a different picture is observed even after annealing.  The nematic order parameter reaches a plateau at long times (see Fig.~\ref{fig3}), with average 
values lower than the corresponding
values in the equilibrium phase at the same temperature ($S=0.878, 0.920$ in the quenched system,  $S=0.964, 0.962$ in the equilibrium system for $T^{*}=0.2, 0.3$ respectively).  The pair distribution functions for a system quenched to  $T^{*}=0.2$ and then annealed for 120,000 iterations are shown in Fig.~\ref{fig4}. There is no smectic layer formation
and no translational order perpendicular to the director, i.e. structurally the quenched system is nematic-like. 

We now show that the system is frozen in this nematic state both translationally and orientationally. The dynamical properties of the system are characterized by
the translational diffusion coefficients and orientational relaxation times. Diffusion 
coefficients parallel and perpendicular to the director are evaluated from mean-squared displacements as follows:
\begin{equation}
D_{\parallel} = \lim_{t \rightarrow \infty} \frac{1}{6 t} <(r_{\parallel}(t+t_{0})-r_{\parallel}(t_{0}))^{2}>
\end{equation}
\begin{equation}
D_{\perp} = \lim_{t \rightarrow \infty} \frac{1}{6 t} <({\bf r}_{\perp}(t+t_{0})-{\bf r}_{\perp}(t_{0}))^{2}>.
\end{equation}
 Both diffusion constants in the glassy phase 
are very small, about 100--1000 times smaller than corresponding values in the nematic phase which are of order 0.1 in dimensionless units.  However, the mean-squared displacement remains a monotonic function of time, unlike the crystalline phase where the mean-squared displacement oscillates about the equilibrium positions. The ratio of the two diffusion 
constants $\frac{D_{\parallel}}{D_{\perp}}$ is approximately 4 in the glass phase, comparable to values in the nematic phase, and very different from the equilibrium values in the smectic and crystalline phases where the ratio is less than one and approximately equal to one respectively. 

To show that the system freezes orientationally we use the spin-glass analogy and compute an Edwards-Anderson type \cite{EA} of  correlation function:
\begin{equation}
C(t) = \frac{1}{N} \sum_{i=1}^N \{ <{\bf u}_{i}(t+t_{0}){\bf u}_{i}(t_{0})> - <{\bf u}_{i}(t+t_{0})><{\bf u}_{i}(t_{0})> \}.
\label{C}
\end{equation}
 The orientational relaxation time is given by the slope of the logarithmic plot of $C(t)$
versus time $t$ and is  approximately 2 orders of
magnitude larger in the glass phase compared with the nematic phase,
as shown in Table \ref{table1}. Thus, our evidence suggests that a glass phase has formed at this temperature with simultaneous freezing of translational and orientational degrees of freedom.

Finally we consider the effects of different quench rates. If we quench the system instantaneously from $T^{*} = 1.2$ to $T^{*} =0.2$ we obtain a glass. However, if the glass is then annealed for several thousand time steps  crystallization occurs. This shows that the
glass becomes more unstable with increasing cooling rate, as seen in simulations of isotropic Lennard-Jones systems \cite{Nose}. On the other hand, with a ``slow''cooling rate corresponding to a temperature reduction of $\Delta T^{*}= 0.1$ every 1000 iterations we find a stable glassy phase. However, the slower quench yields a different structure than the one shown in Fig.~\ref{fig4} because of the smectic phase that intervenes between the nematic and crystalline phases under equilibrium conditions.   At the slower cooling rate the system
starts to form layers typical of the smectic phase, but does not crystallize. There
are no correlations between the different layers and order within the layers is 
not perfect. These effects can be seen in Fig.~\ref{fig5} where we show the positions of the molecules in three adjacent layers of the MD cell when viewed from above for the ``slowly'' quenched system at $T^{*}= 0.2$. Fig.~\ref{fig6} shows a similar view in a system which is allowed to reach thermal equilibrium at the same temperature and a crystalline phase is formed.

\acknowledgements
We are grateful to Dr. George Loriot for helpful discussions. This work was supported  by the National Science
Foundation under grants no. DMR-9217290 and DMR-9528092. Some of the computational work was performed at the Theoretical Physics Computing Facility of Brown University.

\begin{figure}
\caption{The potential energy in dimensionless units as a function of annealing time (also in dimensionless units) for a Gay-Berne fluid quenched from $T^{*}=1.2$ to the temperatures indicated. The quench rate corresponded to a reduction in temperature of 0.1 every 100 iterations.  Note the absence of a break in the curves corresponding to the temperatures $T^{*}=0.2$ and 0.3, unlike the two higher temperatures. }
\label{fig1}
\end{figure}

\begin{figure}
\caption{The pair distribution function versus distance in dimensionless units ($r^{*}\equiv r/\sigma_o$) after a quench to $T^{*}=0.4$ with the same quench rate as in Fig.\ \protect\ref{fig1}. The distribution functions were evaluated after an annealing run of 120,000 iterations, and averaged over the subsequent 120,000 iterations.  The solid curve is the orientationally averaged function. The dashed curve is the distribution function parallel to the director (averaged over the perpendicular direction) plotted against distance parallel to the director. The dashed-dotted curve is the distribution function perpendicular to the director (averaged over the parallel direction) plotted against distance perpendicular to the director. }
\label{fig2}
\end{figure}

\begin{figure}
\caption{The nematic order parameter as a function of annealing time for quenches to the temperatures indicated. The quench rate is the same as in Fig.\ \protect\ref{fig1}.  Note the lower value of the order parameter for the lowest of four temperatures shown.}
\label{fig3}
\end{figure}

\begin{figure}
\caption{The pair distribution function versus distance in dimensionless units after a quench to $T^{*}=0.2$ with the same quench rate as in the previous figures. The distribution functions were evaluated after an annealing time of 120,000 iterations, and averaged over the subsequent 120,000 iterations. The line style of the three curves is as in  Fig.\ \protect\ref{fig2}. Note the absence of smectic order in the parallel distribution function (the lowest curve), and the absence of two well-formed subsidiary peaks in the orientationally-averaged distribution function (the solid curve). Compare with Fig.\ \protect\ref{fig2}.}
\label{fig4}
\end{figure}

\begin{figure}
\caption{The locations of the molecules in the three innermost smectic layers of the MD cell (there are fifteen layers in total) for a system quenched to $T^{*}=0.2$ at the ``slow'' cooling rate and then annealed for 200,000 iterations.  The triangles correspond to molecules in the middle layer of the three layers shown. The boxes and circles correspond to the two layers which flank the middle one. Note the absence of correlations in molecular positions between layers, and the imperfect hexagonal order within each layer. Compare with Fig.\ \protect\ref{fig6} which shows a system at the same temperature in thermal equilibrium.}
\label{fig5}
\end{figure}

\begin{figure}
\caption{The locations of the molecules in the three innermost smectic layers of the MD cell for a system in thermal equilibrium at $T^{*}=0.2$. The notation used is the same as in Fig.\ \protect\ref{fig5}. In this run, which produced an equilibrated crystalline phase, the potential was truncated perpendicular to the director at a distance of 2.5. Two side-by-side molecules at that separation have the same potential energy as two end-to-end molecules separated by the spherical cutoff distance of 3.8.  }
\label{fig6}
\end{figure}
\begin{table}
\caption{The orientational relaxation time constant $\tau$ and the nematic order parameter $S$  at several representative temperatures for a nematic in thermal equilibrium and a glass quenched at two different rates.  The fast quench rate corresponds to a temperature reduction of 0.1 every 100 iterations and the slow rate to the same reduction every 1000 iterations. The glass formed at the faster quench rate has less nematic order and significantly longer relaxation times.}
\begin{tabular}{ccc}
&$\tau$&$S$\\
\tableline
Nematic, $T^{*}=1.0$&$5.1 \pm 0.2$&$0.84 \pm 0.01$\\
Nematic, $T^{*}=1.1$&$5.2 \pm 0.2 $&$0.79 \pm 0.01$\\
Glass, slow quench, $T^{*}=0.2$&$394 \pm 6$&$0.888 \pm 0.004$\\
Glass, slow quench, $T^{*}=0.3$&$152 \pm 1$&$0.904 \pm 0.007$\\
Glass, slow quench, $T^{*}=0.4$&$48\pm 2$&$0.969 \pm 0.002$\\
Glass, fast quench, $T^{*}=0.2$&$676 \pm 9$&$0.878 \pm 0.004$\\
Glass, fast quench, $T^{*}=0.3$&$93 \pm 1$&$0.920 \pm 0.004$\\
Glass, fast quench, $T^{*}=0.4$&$100 \pm 2$&$0.961 \pm 0.002$\\
\end{tabular}
\label{table1}
\end{table}


\begin{references}

\bibitem{expt} For experimental studies see, e.g., J. I. Spielberg and E. Gelerinter, Phys. Rev. B{\bf 32}, 3647 (1984), L. Rosta, Mol. Cryst. Liq. Cryst, {\bf 127}, 195 (1985), V. K. Dolganov, R. Fouret, C. Gors, and M. More, Phys. Rev. E {\bf 49}, 5230 (1994).

\bibitem{SG} See, e.g., J. Villain, Z. Phys. B{\bf33}, 31, (1979); M. Gabay and G. Toulouse, Phys. Rev. Lett. {\bf 47}, 201 (1981).

\bibitem{GB} J. G. Gay and B. J. Berne, J. Chem. Phys. {\bf 74}, 3316 (1981).

\bibitem{deM} E. DeMiguel, L. F. Rull, M. K. Chalam, and K. E. Gubbins, Mol.
Phys. {\bf 74}, 405 (1991).

\bibitem{Luck90} G. R. Luckhurst, R. A. Stephens, and R. W. Phippen, Liq.
Cryst. {\bf 8}, 451 (1990).

\bibitem{LS93} G. R. Luckhurst and P. S. J. Simmonds, Mol. Phys. {\bf 80}, 233
(1993).

\bibitem{DeRull} E. DeMiguel, L. F. Rull, and K. E. Gubbins, Phys. Rev.
A{\bf45}, 3813 (1992).

\bibitem{Sarman1}S. Sarman and D. J. Evans, J. Chem. Phys. {\bf 99}, 620
(1993); S. Sarman, J. Chem. Phys. {\bf 101}, 480 (1994).

\bibitem{Sarman2} S. Sarman and D. J. Evans, J. Chem. Phys. {\bf 99}, 9021
(1993).

\bibitem{Smondyrev} A. M. Smondyrev, G. B. Loriot and R. A. Pelcovits, Phys. Rev. Lett. {\bf 75}, 2340 (1995).

\bibitem{NH} S. Nose, Mol. Phys. {\bf 52}, 255 (1984); W. G. Hoover, Phys. Rev.
A{\bf 31}, 1695 (1985); S. Nose, Mol. Phys. {\bf 50}, 1055 (1983).

\bibitem{Pressure} At this dimensionless pressure the Gay-Berne fluid has isotropic, nematic, smectic and crystalline phases; see ref. \cite{deM}. 

\bibitem{Dolganov} V. K. Dolganov, N. Kroo, L. Rosta, E. F. Sheka and J. Szabon, Mol. Cryst. Liq. Cryst. {\bf 127}, 187 (1985).

\bibitem{EA} S. F. Edwards and P. W. Anderson, J. Phys. F{\bf 5}, 965 (1975).

\bibitem{Nose} S. Nose and F. Yonezawa, Sol. State Comm. {\bf 56}, 1005 (1985).

\end{references}
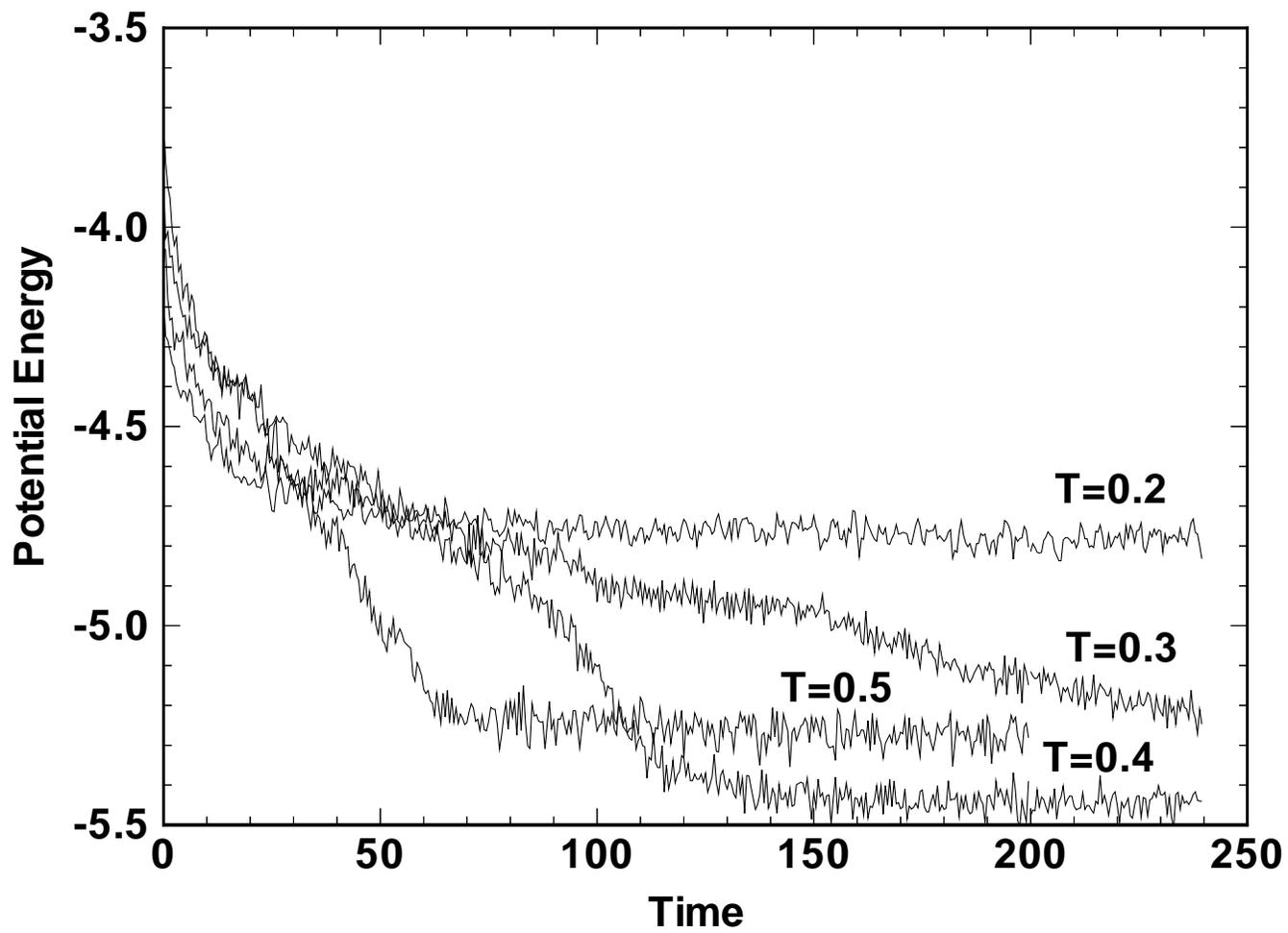
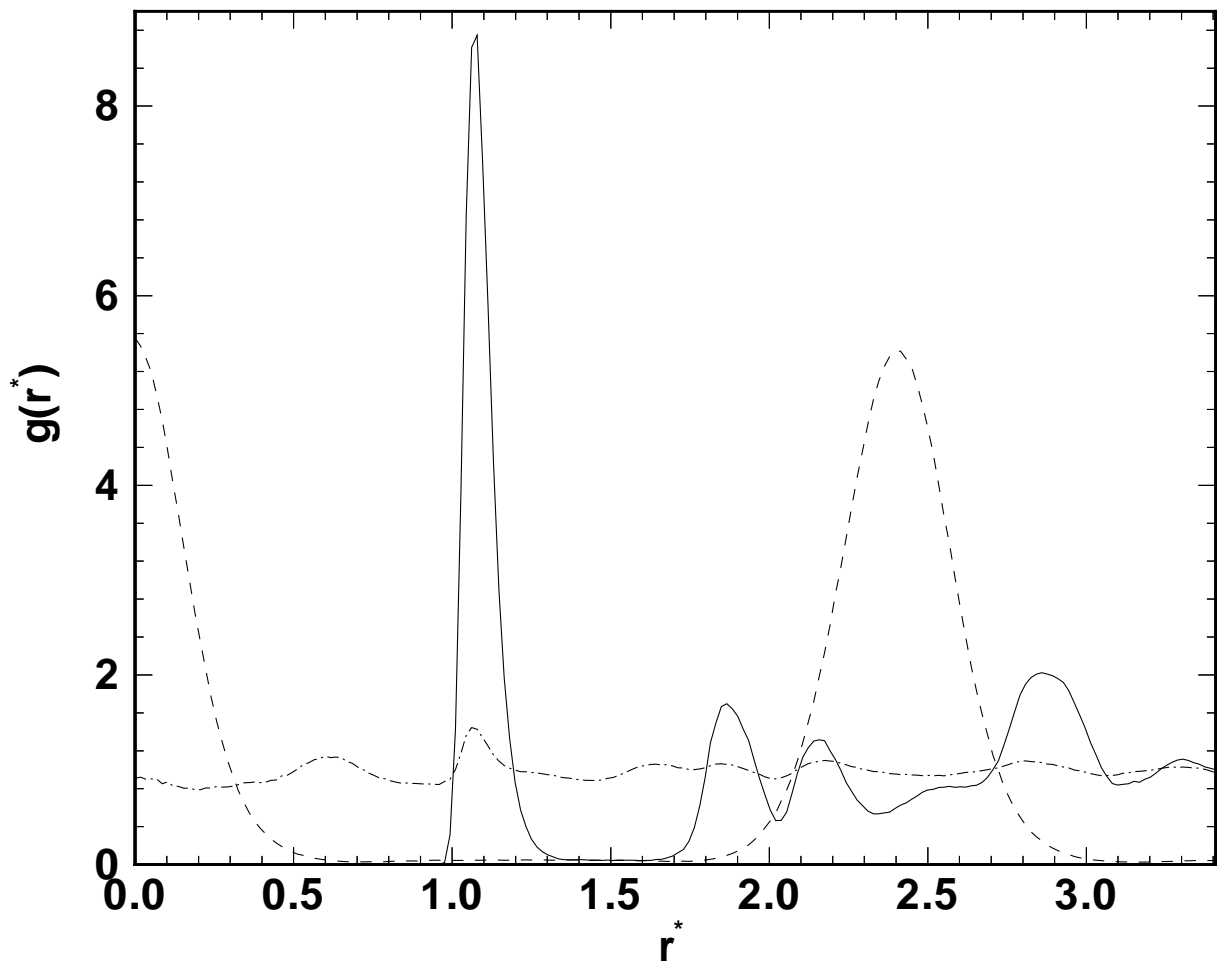
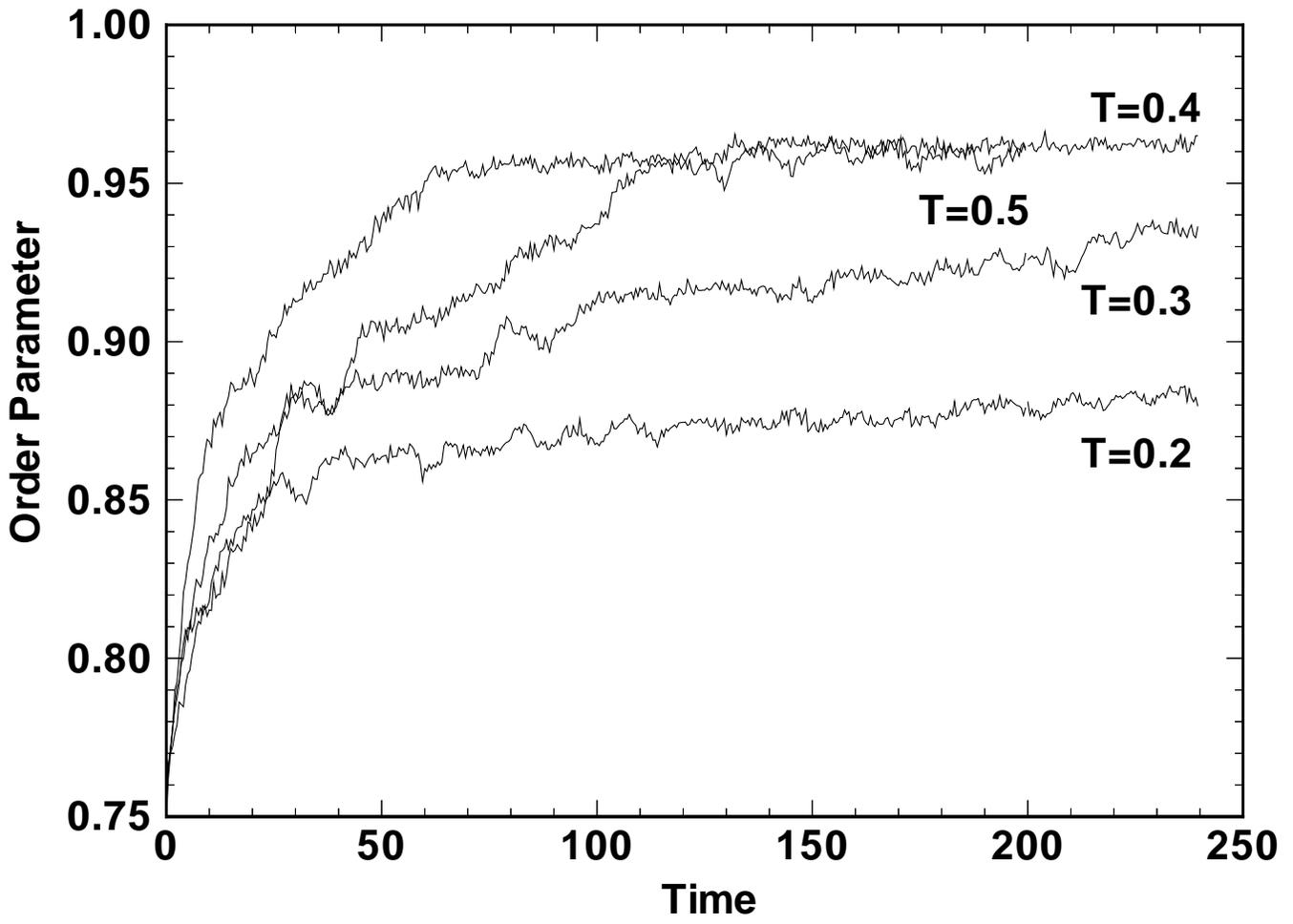
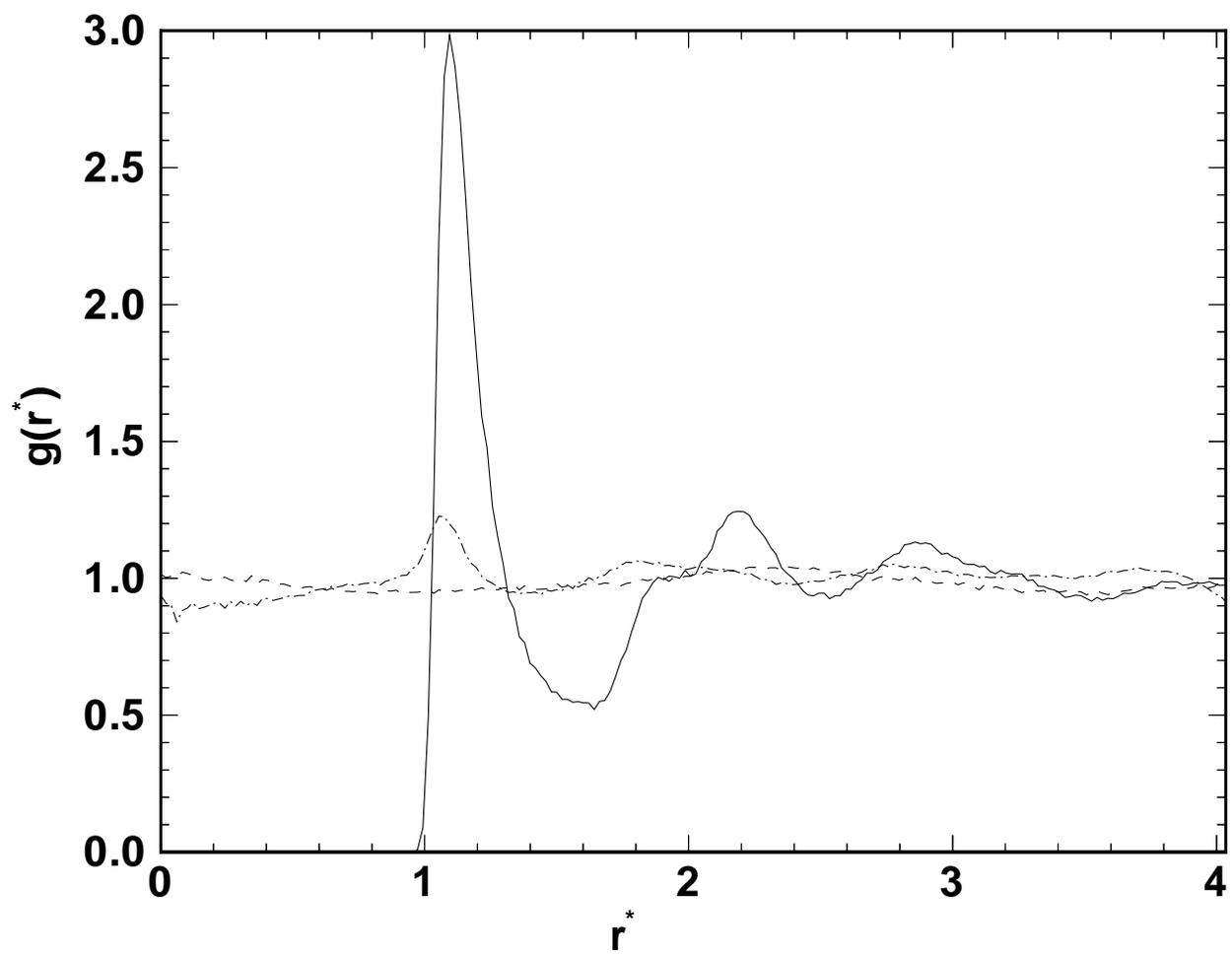
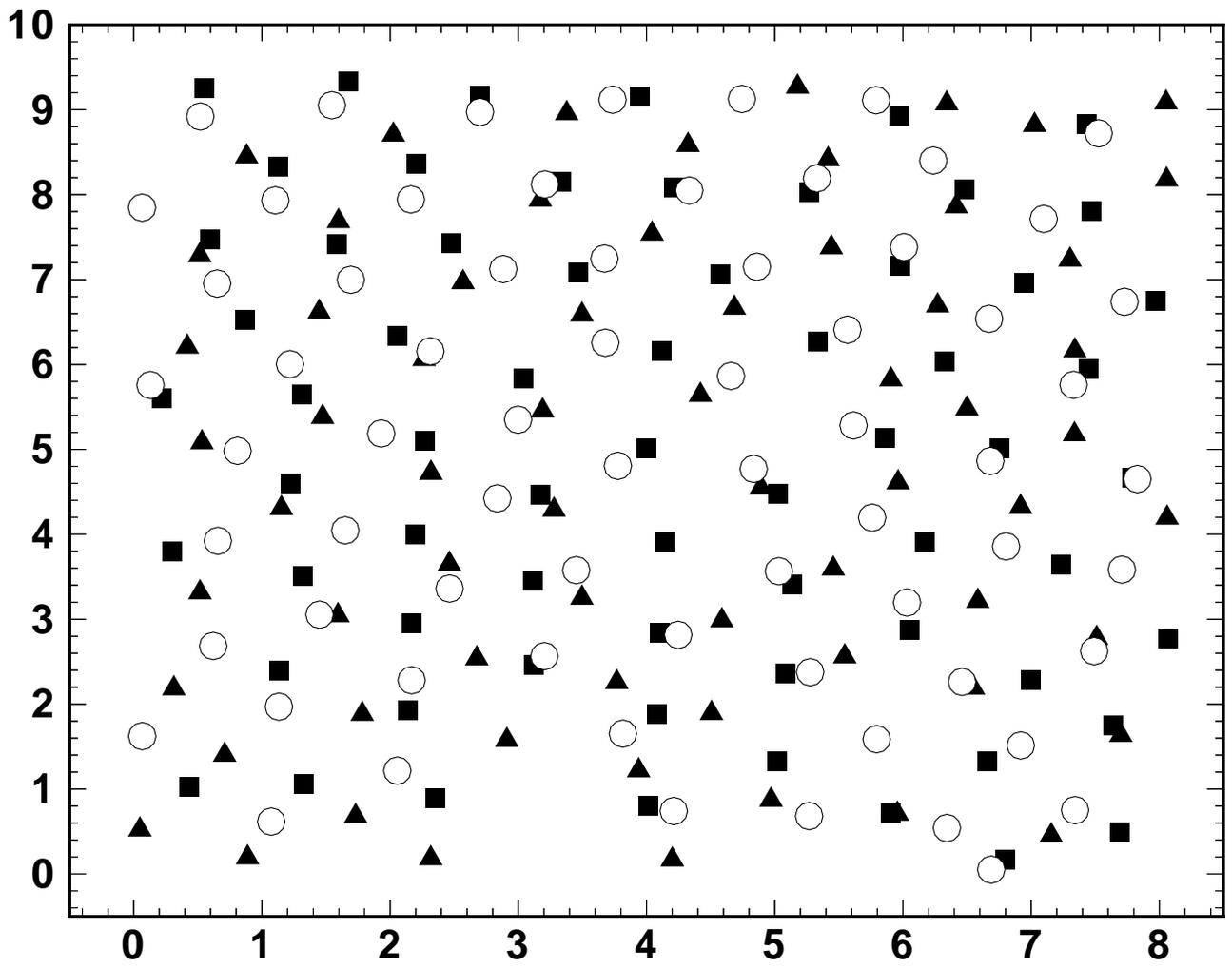
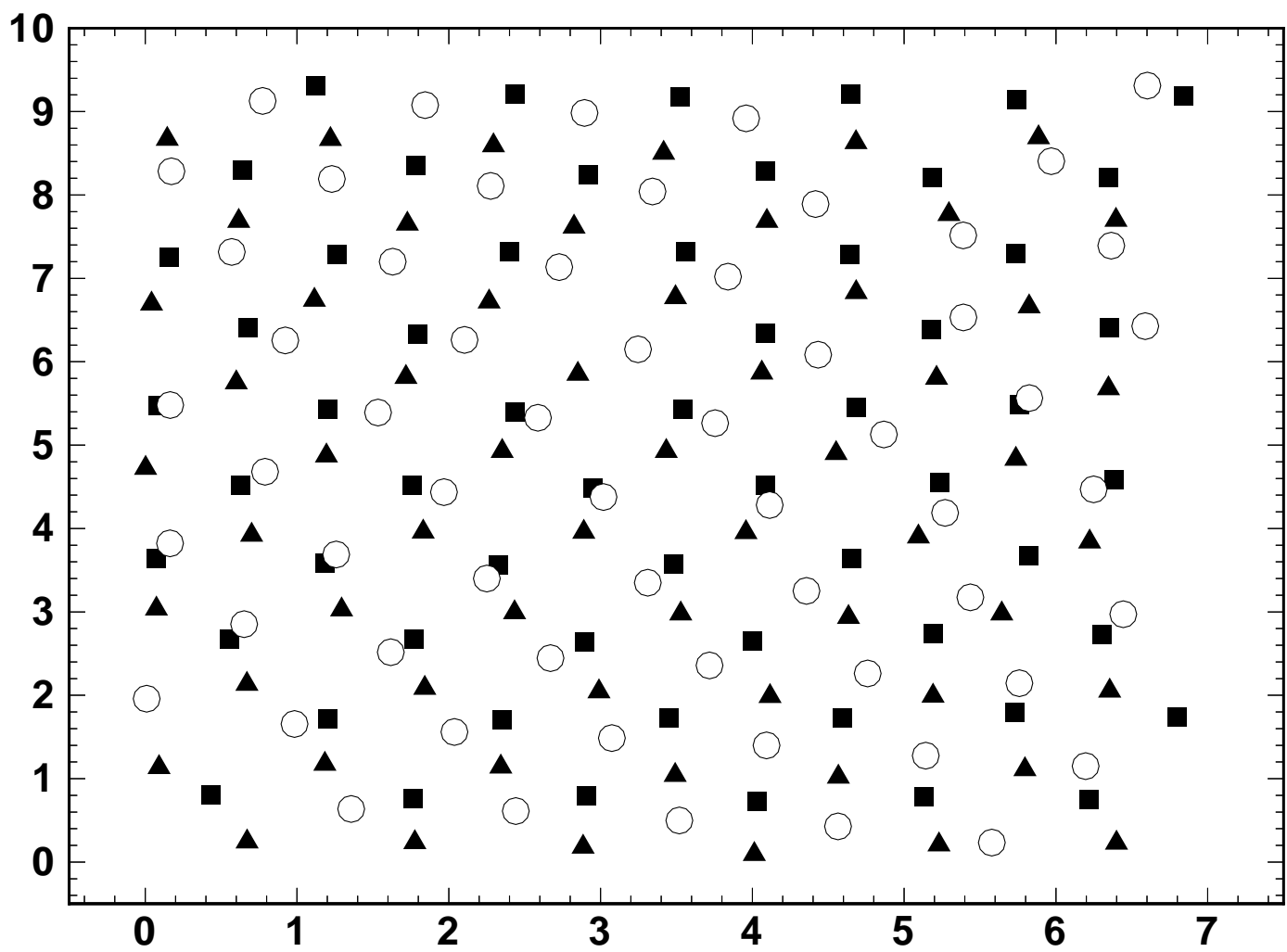
\end{document}